\documentclass{ws-procs975x65}            
\begin{document}
\title{Hydrodynamics in $1+1$ dimensions from Maxwell-Chern-Simons theory in $AdS_3$}

\author{Han-Chih Chang$^a$, Mitsutoshi Fujita$^{b}$\footnote{Speaker. E-mail: mitsutoshi.fujita@yukawa.kyoto-u.ac.jp}, Matthias Kaminski$^c$}

\address{$^a$ Department of Physics
University of Virginia
382 McCormick Rd.
PO Box 400714
Charlottesville, VA, USA  \\
$^b$ Yukawa Institute for Theoretical Physics, Kyoto University, Kyoto, Japan \\ 
$^c$Department of Physics and Astronomy, University of Alabama, Tuscaloosa, AL 35487, USA}

\begin{abstract}
In this presentation we review our work on Abelian Maxwell-Chern-Simons theory in three-dimensional $AdS$ black brane backgrounds, with both integer and non-integer Chern-Simons coupling. Such theories can be derived from several string theory constructions,
and we found exact solutions in the low frequency, low momentum limit ($\omega, k \ll T$,~the hydrodynamic limit). Our results are translated into correlation functions of  vector operators in the dual strongly coupled $1+1$-dimensional quantum 
field theory with a chiral anomaly at non-zero temperature $T$, via the holographic correspondence. 
The applicability of the hydrodynamic limit is discussed, together with the comparison between an exact field theoretic computation and the found holographic correlation functions in the conformal case.
\end{abstract}


\bodymatter

\section{Introduction}
The hydrodynamic behavior of the $1+1$ dimensional theory dual to the $1+2$-dimensional Maxwell-Chern-Simons (MCS) theory on the $AdS_{3}$ black brane has not been well-understood with the presence of the massive sector. In our work~\cite{Chang:2014jna}, 
we specifically focus on this hydrodynamic limit, that is when temperature is much larger than frequency and momentum.
Related work on the Maxwell-Chern-Simons theory includes analysis in the zero temperature limit~\cite{Minces:1999tp,Andrade:2011sx}, and in the hydrodynamic limit with the flat sector~\cite{Jain:2012rh,Loganayagam:2011mu,Jensen:2010em}. 
Pure Yang-Mills theory (without Chern-Simons term) in the $AdS_{3}$ black brane background is also interesting, as it provides an example for a $1+1$ dimensional holographic superconductor at large charge density by virtue of the large $N$ limit~\cite{Gao:2012yw}. It is known that only the special (Neumann) boundary condition~\cite{Marolf:2006nd} is allowed for the gauge field in such a theory. 

It was suggested that the Maxwell-Chern-Simons theory was dual to the chiral Luttinger liquid, namely, a theory of electrons in the gapless edge state in the context of the fractional quantum Hall effect~\cite{Fujita:2009kw}. In the chiral Luttinger theory, excitations travel into one particular direction, which is the drift motion in the presence of a strong magnetic field. In this presentation, our holographic model of the probe Maxwell-Chern-Simons action is conjectured to be dual to a chiral Luttinger liquid coupled to $1+1$-dimensional thermal CFT ($AdS_{3}$ black brane background)~\cite{Chang:2014jna}. 

\section{1+1 dimensional anomalous hydrodynamics}
In order to compare with the gravity dual, on the field theory side we consider the leading terms of the hydrodynamic expansion in the presence of a anomaly
\begin{eqnarray}\label{Eq1}
&T^{\mu\nu}=\epsilon_{I} u^{\mu}u^{\nu}+P_{I}\Delta ^{\mu\nu}+\dots,\quad  J^{\mu}=\rho_{I} u^{\mu}+\chi \mu \epsilon^{\mu\nu}u_{\nu}+\dots,\\
&D_{\mu}J^{\mu}=-\dfrac{\chi}{2}\epsilon^{\mu\nu}F_{\mu\nu},\quad D_{\mu}T^{\mu\nu}=J_{\mu}F^{\nu\mu}, \label{Eq2}
\end{eqnarray}
where $u^{\mu}$ denotes the fluid velocity ($u_{\mu}u^{\mu}=-1$), and $\Delta^{\mu\nu}=g^{\mu\nu}+u^{\mu}u^{\nu}$. Here, $\epsilon_{I},\ P_{I},\ \rho_{I}$ are the energy density, the pressure, and the charge density in equilibrium, respectively. In the second line, $D_{\mu}$ is the covariant derivative and $-\chi \epsilon^{\mu\nu}F_{\mu\nu}/2$ is the anomaly term associated with the anomaly in the global axial $U(1)$ symmetry.
The terms in \eref{Eq1} are the leading contributions to the conserved currents $T^{\mu\nu}$ and $J^\mu$ in the hydrodynamic limit~\cite{Jain:2012rh,Loganayagam:2011mu}. Note that it has been shown that the anomalous transport terms do not lead to any entropy production~\cite{Loganayagam:2011mu}, and
all the terms in \eref{Eq1} are non-dissipative.
Transport peaks (associated with dissipation) are not observed in the holonomy current (dual to a pure Chern-Simons theory), because the holomorphic decomposition does not mix the two sectors of different chirality. We are going to argue that the Maxwell-Chern-Simons theory is holographically dual to a field theory whose low-energy limit is described by anomalous hydrodynamics including terms with the parity breaking epsilon symbol in \eref{Eq1} and (\ref{Eq2}). Note also that temperature does not break conformal invariance in $1+1$ dimensions.

\section{Analysis of the gravity dual: Maxwell-Chern-Simons action}
On the gravity side, the Maxwell-Chern-Simons theory is analyzed on the $AdS_{3}$ black brane background. The $\theta$ Maxwell-Chern-Simons action ($\theta-$MCS) is given by~\cite{MCS1}
\begin{equation}\label{Action}
S=T\int d^{3}x(\sqrt{-g}f_{\mu\nu}f^{\mu\nu}+\theta \epsilon^{\mu\nu\rho}a_{\mu}f_{\nu\rho}),
\end{equation}
where $T\theta$ is the Chern-Simons level and $T$ is a tension.
There are many string theory embeddings of the Maxwell-Chern-Simons theory, yielding both integer and non-integer $\theta$~\cite{Karndumri:2013dca}. A rather famous one is the dimensional reduction of a type IIB string theory embedding on $AdS_{3}\times S^{3}\times M_{4}$ ($M_{4}=S^{3}\times S^{1}$ or $T^{4}$)~\cite{Gukov:2004ym}. The resulting Maxwell-Chern-Simons theory has two Chern-Simons terms with levels of opposite sign. Moreover, a string theory reduction on $AdS_{3}\times S^{3}\times T^{4}$ yields $\theta =2-$MCS, and leads to the presence of a scalar potential in addition to the cosmological constant~\cite{Detournay:2012dz}. Maxwell-Chern-Simons theories can also be obtained from the probe brane actions on the supergravity backgrounds. More specifically, a Chern-Simons term arises from the Wess-Zumino term of the probe brane actions. The D3/D7 system with $\sharp$ND$=8$, which has chiral fermions as zero modes from $3-7$ strings, leads to $\theta =-4-$MCS~\cite{Harvey:2008zz}.

To derive the renormalized retarded Green's function, counter-terms and boundary terms need to be added to the action \eref{Action} as follows~\cite{deHaro:2000xn,Yee:2011yn}: 
\begin{eqnarray}\label{AOF}
 &S_{reg}=S+
 T\int d^2x\sqrt{-\gamma}\Big(
 2 K_0 a_i f^i +
 K_1f_{ij}f^{ij}+
 K_3f_if^i+2K_0'f^i \Delta a_i \nonumber \\
 & +K_3'f_i\Delta f^i 
 + K_p a_ia^i
 + R_1 (n^\mu\partial_\mu f_{ij})^2
 + R_2 (n^\mu\partial_\mu f_{i})^2
 + R_3 (n^\mu\partial_\mu a_i)^2\nonumber \\
& + Q_1 \gamma^{ij} a_i\Delta a_j 
 + Q_2 \gamma^{ii'}\gamma^{jj'} f_{ij} \Delta f_{i'j'}
 + Q_3 \gamma^{ij} a_i \Delta f_j
 \Big)\nonumber \\
& +T \int [
 K_2\epsilon^{ij}f_ia_j
 + K_2'\epsilon^{ij} a_j \Delta f_i
 + Q_4 \epsilon^{ij} a_i \Delta a_j
 + Q_5 \epsilon^{ij} f_i \Delta f_j\nonumber \\
& + Q_6 \epsilon^{ij} f_i \Delta a_j
 ]+ \mathcal{O}(\log u)+\dots, 
\end{eqnarray}
where $i,j=1,2$, $f^{i}=f^{\mu i}n_{\mu}$, 
$n_{\mu}$ is the vector normal  to the boundary satisfying $n_{\mu}n^{\mu}=1$, 
$\epsilon^{ij}$ is the Levi-Civita symbol, and $\Delta =\gamma^{ij}\partial_{i}\partial_{j}$ is the boundary Laplacian in terms of the induced boundary metric $\gamma$. 
Note that counter-terms multiplied by $\mathcal{O}(\log)$ terms are to be included, although not explicitly displayed here. Assuming $\theta >0$, solutions of the gauge field are expanded near the $AdS$ boundary ($u\sim 0$)
\begin{eqnarray}
a_t & = & B_t^{(-\theta)} u^{-\theta} +\dots+ a_t^{(0)} +\dots + B_t^{(\theta)} u^{\theta} + \log(u) \left (
B_t^{(0),\,\text{log}} + B_t^{(1),\,\text{log}} u+ B_t^{(2),\,\text{log}} u^2 + \dots
\right )
 \, , \nonumber\\
a_x & = & B_x^{(-\theta)} u^{-\theta} +\dots+ a_x^{(0)} + \dots + B_x^{(\theta)} u^{\theta} + \log(u) \left (
B_x^{(0),\,\text{log}} + B_x^{(1),\,\text{log}} u+ B_x^{(2),\,\text{log}} u^2 + \dots
\right )
 \, . \nonumber\\
\end{eqnarray}
Imposing the EOM, one constraint implies $B_{t}^{(-\theta)}=B_{x}^{(-\theta)}$, where $B_{t}^{(-\theta)}$ is the source term and $B_{t}^{(\theta)},\ B_{x}^{(\theta)}$ are VEV terms of the vector operator $\langle \mathcal{O}_{1} \rangle$. The VEV terms  are related with the source terms via the IR boundary conditions. $a_{i}^{(0)}$ is dual to the chiral current $\langle \mathcal{O}_{\pm}\rangle $, and it is a flat connection satisfying $da^{(0)}=0$.

 Let's focus on the case of $\theta =2$-MCS which corresponds to the string theory embedding on $AdS_{3}\times S^{3}\times M_{4}$. 
The on-shell action is holographically renormalized and becomes finite, with the following choice of counter-term coefficients:
\begin{eqnarray} \label{eq6}
&K_{0}=1,\quad K_{1}=-\dfrac{1}{4},\quad K_{2}=0,\quad K_{2}'=2K_{0}'+\dfrac{1}{2}-Q_{1},\quad K_{3}=\dfrac{1}{4},\nonumber \\
& K_{3l}'=\dfrac{1}{2},\quad K_{p}=0,\quad R_{1}=\dfrac{1}{32},\quad R_{2}=-\dfrac{1}{16}.
\end{eqnarray}
Then the renormalized Green's functions are derived in the hydrodynamic limit by performing the variation of the finite on-shell action, yielding the vector operator correlators and chiral current correlator
\begin{eqnarray}
&\langle \mathcal{O}_1 \mathcal{O}_1 \rangle = \dfrac{\delta^2 S_\text{reg}}{\delta B_t^{(-2)}\delta B_t^{(-2)}} = T i 4 \omega + \mathcal{O} (2),
 \, \\
 &\langle \mathcal{O}_{\pm} \mathcal{O}_{\pm} \rangle = \dfrac{\delta^2 S_\text{reg}}{\delta a_{\mp}^{(0)}\delta a_{\mp}^{(0)}} =\pm 8T\dfrac{q_{\pm}}{q_{\mp}} \, , \label{FLA1}
\end{eqnarray}
with $a_\pm^{(0)} = (a_x^{(0)}\pm a_t^{(0)})/\sqrt{2}$, where $q_\pm \propto (k\pm\omega)$. 
It is found that the hydrodynamic two-point functions exactly agree with the appropriate CFT correlation function at the leading order. Note that the scaling behavior of temperature is omitted and can be recovered by dimensional analysis.

\section{Summary}
We have outlined our obtained results~\cite{Chang:2014jna}: First, hydrodynamic solutions of MCS theory were obtained on $AdS_{3}$ black branes for $\theta =-2,-4,\dots$, implying that analytic solutions exist for all even Chern-Simons levels. 
Second, the on-shell action of the MCS theory was holographically renormalized for $\theta =2$ (and non-integer $\theta$), satisfying a well-defined variational principle. Details in our field theory calculation of the current correlators can be found in our original publication~\cite{Chang:2014jna}.

\bibliographystyle{ws-procs975x65}
\bibliography{ws-pro-sample}

\begin{thebibliography}{11}
\bibitem{Minces:1999tp}
P.~Minces and V.~O. Rivelles, 
    {\em Phys.Lett.} {\bf B455} (1999) 147--154.

\bibitem{Andrade:2011sx}
T.~Andrade, J.~I. Jottar, and R.~G. Leigh, 
   {\em JHEP}{\bf 1205} (2012) 071.

\bibitem{Jain:2012rh}
S.~Jain and T.~Sharma, 
   {{\tt
  arXiv:1203.5308}}.

\bibitem{Loganayagam:2011mu}
R.~Loganayagam, 
 {{\tt arXiv:1106.0277}}.

\bibitem{Jensen:2010em}
K.~Jensen, 
 {\em JHEP}
  {\bf 1101} (2011) 109, {{\tt
  arXiv:1012.4831}}.

\bibitem{Gao:2012yw}
X.~Gao, M.~Kaminski, H.-B. Zeng, and H.-Q. Zhang, 
  {{\tt arXiv:1204.3103}}.


\bibitem{Marolf:2006nd}
D.~Marolf and S.~F. Ross, 
   {\em JHEP} {\bf 11} (2006) 085,
  {{\tt hep-th/0606113}}.

\bibitem{Fujita:2009kw}
M.~Fujita, W.~Li, S.~Ryu, and T.~Takayanagi, 
    {\em
  JHEP} {\bf 0906} (2009) 066, {{\tt  arXiv:0901.0924}}.

\bibitem{Chang:2014jna} 
  H.~C.~Chang, M.~Fujita and M.~Kaminski,
  {\em JHEP} {\bf 1410}, 118 (2014)
  {\tt arXiv:1403.5263}.

\bibitem{MCS1}
S. Deser,  R. Jackiw, S. Templeton,  {\em Annals Phys.} {\bf 140} (1982) 372-411;  {\bf 281} (2000) 409-449 

\bibitem{Gukov:2004ym}
S.~Gukov,~E.~Martinec,~G.~W.~Moore,~and~A.~Strominger, 
{\em Adv.Theor.Math.Phys.} {\bf 9} (2005) 435--525,
  {{\tt hep-th/0403090}}.

\bibitem{Karndumri:2013dca}
P.~Karndumri and E.~O. Colg{\' a}in, 
   {\em JHEP} {\bf 1310} (2013) 094,
  {{\tt arXiv:1307.2086}}.

\bibitem{Detournay:2012dz}
S.~Detournay and M.~Guica, 
  {\em JHEP}
  {\bf 1308} (2013) 121, {\tt  arXiv:1212.6792}.

\bibitem{Harvey:2008zz}
J.~A. Harvey and A.~B. Royston, 
    {\em JHEP} {\bf 0808} (2008) 006,
  {{\tt arXiv:0804.2854}}.
  
  \bibitem{deHaro:2000xn}
S.~de~Haro, S.~N. Solodukhin, and K.~Skenderis, 
    {\em Commun.Math.Phys.} {\bf 217} (2001) 595--622,
  {{\tt hep-th/0002230}}.

\bibitem{Yee:2011yn}
H.-U. Yee and I.~Zahed, 
    {\em JHEP} {\bf 1107} (2011) 033,
  {{\tt arXiv:1103.6286}}.


\end{thebibliography}

\end{document}